\def\kms{\ifmmode {\rm\,km\,s^{-1}}\else
    ${\rm\,km\,s^{-1}}$\fi}
\def\ms{\ifmmode {\rm\,m\,s^{-1}}\else
    ${\rm\,m\,s^{-1}}$\fi}
\def\kmsMpc{\ifmmode {\rm\,km\,s^{-1}\,Mpc^{-1}}\else
    ${\rm\,km\,s^{-1}\,Mpc^{-1}}$\fi}
\def\kpc{{\rm\,kpc}}

\def\msun{\ifmmode {\rm\,M_\odot}\else ${\rm\,M_\odot}$\fi}
\def\Msun{\ifmmode {\rm\,M_\odot}\else ${\rm\,M_\odot}$\fi}
\def\lsun{\ifmmode {\rm\,L_\odot}\else ${\rm\,L_\odot}$\fi}
\def\Lsun{\ifmmode {\rm\,L_\odot}\else ${\rm\,L_\odot}$\fi}
\def\rsun{\ifmmode {\rm\,R_\odot}\else ${\rm\,R_\odot}$\fi}
\def\Rsun{\ifmmode {\rm\,R_\odot}\else ${\rm\,R_\odot}$\fi}
\def\pc{{\rm\,pc}}

\def\cmtw{\ifmmode {\rm\,cm^{-2}}\else ${\rm\,cm^{-2}}$\fi}
\def\cmthr{\ifmmode {\rm\,cm^{-3}}\else ${\rm\,cm^{-3}}$\fi}
\def\yr{{\rm\,yr}}
\def\gyr{{\rm\,Gyr}}
\def\Gyr{{\rm\,Gyr}}

\def\ergps{\ifmmode {\rm\,erg\,s^{-1}}\else ${\rm\,erg\,s^{-1}}$\fi}
\def\ergpscm2{\ifmmode {\rm\,erg\,s^{-1}\,cm^{-2}}\else
\def{\rm\,erg\,s^{-1}\,cm^{-2}}$\fi}

\def\eg{{\it e.g.}}
\def\deg{\ifmmode {^{\circ}}\else {$^\circ$}\fi}
\def\degr{\ifmmode {^{\circ}}\else {$^\circ$}\fi}
\def\degs{\ifmmode {^{\circ}}\else {$^\circ$}\fi}

\def\etal{{\it et al.~}}
\def\hkpc{h^{-1}{\rm kpc}}
\def\hsfMpc{h_{75}^{-1}{\rm Mpc}}
\def\hMpc{h^{-1}{\rm Mpc}}
\def\hthrMpc{h^{-3}{\rm Mpc}^3}
\def\Ho{\ifmmode {\rm\,H_\circ}\else ${\rm\,H_\circ}$\fi}
\def\hnot{\ifmmode {\rm\,H_\circ}\else ${\rm\,H_\circ}$\fi}
\def\h0{\ifmmode {\rm\,H_\circ}\else ${\rm\,H_\circ}$\fi}
\def\hnotunit{\ifmmode {\rm\,km\,s^{-1}\,Mpc^{-1}}\else
    ${\rm\,km\,s^{-1}\,Mpc^{-1}}$\fi}
\def\Lya{Lyman-$\alpha$}
\def\qnot{\ifmmode {\rm\,q_\circ}\else ${\rm q_\circ}$\fi}
\def\q0{\ifmmode {\rm\,q_\circ}\else ${\rm q_\circ}$\fi}

\def\arcsec{\ifmmode {^{\prime\prime}}\else $^{\prime\prime}$\fi}
\def\asec{\ifmmode {^{\prime\prime}}\else $^{\prime\prime}$\fi}
\def\arcmin{\ifmmode {^{\prime}}\else $^{\prime}$\fi}
\def\amin{\ifmmode {^{\prime}}\else $^{\prime}$\fi}

\documentstyle[emulateapj]{article}
\lefthead{C. V. Manning}
\righthead{Globular Cluster Formation}

\begin{document} 
\title{The Accretion of Lyman Alpha Clouds onto Gas--Rich
Protogalaxies; \\
  A Scenario for the Formation of Globular Star Clusters}
\author{Curtis V. Manning}
\affil{Department of Astronomy, University of California,
	Berkeley, CA  94720 e-mail:cmanning@astro.berkeley.edu}

\begin{abstract}

A satisfactory theory for the formation of globular star clusters
(GCs) has long been elusive, perhaps because their true progenitors
had not yet been guessed.  In this paper I propose a causal
relationship between the strongly decreasing densities of Lyman
$\alpha$ clouds at high--redshift and the formation of globular
clusters --- namely, that GCs were created by the accretion of Lyman
$\alpha$ clouds onto protogalaxies.  I describe a scenario which
involves an inherently stable and orderly cycling of compression and
cooling in the central cores of these clouds during the extended
period of dissipation in the outer regions of gas--rich protogalaxies,
culminating in a burst of efficient star formation.  I demonstrate
that the comoving density of GCs is comparable to that of Lyman
$\alpha$ clouds at high--redshift, that the energetic requirements for
compression to core GC densities can be met, and that the time--scale for
cooling is within obvious limits imposed by dynamical stability.

This dissipative process requires there to be a large column of
dissipated gas about the attractor in order to form GCs.  In adition,
the energy requirements of compression to GC central densities
requires attractor masses greater than that capable of sustaining
circular velocities of $\sim 40 \kms$.  If this scenario is supported
by numerical simulations, then by implication, the GCs were formed at
modest redshifts of $z \sim 1-3$.  This knowledge could help to break
the troublesome degeneracy between look--back time and redshift.  This
model is consistent with a picture of hierarchical galaxy growth over
time scales of many billions of years.

\end{abstract}


\section{Introduction}

Globular clusters (GCs) are at once objects of interest to students of
the Galaxy and to cosmologists --- tothe former, as probes of galaxy
formation and as dynamically pure objects of curiosity, and to the latter
because of their great age, and for what they may tell us about
conditions in the early Universe.  Attempts to explain them have
included models based on the accretion of protogalactic clumps (Searle
\& Zinn 1978; Searle 1980), two--phase collapse (Fall \& Rees 1985;
Murray \etal 1993), and cloud collisions (Murray \& Lin 1989; 1993;
Harris \& Pudritz 1991; McLaughlin \& Pudritz 1996).  In this paper I
suggest a dissipative scenario in which Lyman $\alpha$ clouds are
viewed as the progenitors of the GCs.

Studies of high--redshift Lyman $\alpha$ clouds (\eg, Lu, Wolfe, \&
Turnshek 1991; Bechtold 1994) show that the high line density of
clouds at $z \approx 3$ decreases rapidly with time.  A major
constituent of the Universe at high--redshift, these clouds are not
likely to have disappeared without a trace.  Binney (1976) first
suggested that the earliest star formation in a collapsing protogalaxy
would be within transient sheets of compressed gas generated by the
collision of infalling gas.  Much of the halo may have been created in
this way.  But the formation of GCs requires a more extraordinary
variant of this accretion process.  If clouds are centrally condensed
(\eg, dark matter (DM) held), this would help them to survive the
infall intact, possibly leading to a super--concentrated burst of star
formation, and hence possibly a GC.  The proposed GC formation
scenario utilizes the kinetic energy of infall to compress
clouds to densities comparable to the central regions of GCs.

Below, in referring to galaxian quantities and galactocentric
distances, I use upper--case letters, and for cloud quantities,
lower--case letters.  When this convention isn't applicable, I will use 
subscripts.

\section{Anecdotal Evidence}

Zaritsky (1995) has shown that there are abundance gradient anomalies,
and asymmetries in the distribution of HI at large galactocentric
radii in $1/3$ to $1/2$ of nearby spiral galaxies.  This was
cautiously interpreted as being the result of the accretion of low
metallicity HI clouds within the last few $\gyr$.  More recently a
measure of lopsidedness and anisotropy has been developed from the
Fourier analysis of the surface brightness distributions of disk
galaxies (Zaritsky \& Rix 1997; Rudnick \& Rix 1998; Jiang \& Binney
1998) which has been found to correlate well with the locations of
enhanced star formation and H I anomalies.  These findings support the
conclusions of Zaritsky's (1995) earlier work.

A source for these apparent accretion events may have been found in a
population of low redshift HI clouds, discovered in Hubble Space
Telescope (HST) spectra of low redshift quasar and active galaxy
spectra (Bahcall \etal 1991; Morris \etal 1991; Morris \etal 1993,
Lanzetta \etal 1995; Stocke \etal 1995; Shull, Stocke \& Penton 1996;
Tripp, Lu, \& Savage 1998), and at redshifts out to $z \approx 0.6$
(Chen \etal 1998).  Many of these clouds appear to be clustered about
luminous ($M_V \lesssim -19$) field galaxies.  Recently, Tripp \etal
(1998), adding data from two QSO sightlines, have summarized recent
developments, showing that of those clouds which are found within $2
h_{75}^{-1} {\rm Mpc}$ of a major galaxy, most have a projected
distance of less than 700 kpc.  However, nearest--galaxy distances are
seen to be greater than $2 h_{75}^{-1} {\rm Mpc}$ for as many as $40
\%$ of the clouds, prompting Tripp \etal (1998) to refer to this
population as ``void clouds''.

In a study of high velocity clouds (HVCs) in the Local Group, Blitz
\etal (1998) note that the internal velocity dispersions of remote 
HVCs are strongly peaked, with $\sigma \approx 20 \pm 6.5 \kms$, which
argues for a generally homogeneous population that is only mildly
subgalactic in its virial temperature.  Their dust--to--gas ratios are
at least a factor of three below that of normal Galactic clouds, and
heavy--element abundances well below solar values.  Wakker \etal
(1998) have found that the ``C'' cloud, the closest HVC, has a
metallicity, $Z=0.07 \pm 0.02 \, Z_{\odot}$, well below what would be
expected for a Galactic fountain or tidal tail.  I therefore assume
that these apparently infalling clouds are indeed representatives of
the extragalactic Lyman $\alpha$ population.

It therefore seems likely that the more clustered cloud population is
a major source of the impactors hypothesized by Zaritsky (1995), and
others.  They apparently share the same physical region occupied by
dwarf galaxies that cluster about dominant field galaxies.  If clouds
share the kinematic characteristics of dwarfs as well, then we can
expect that clouds will cluster about their primary within velocities
of $ \leq 400$ km/s (Zaritsky \etal 1997, and references therein).
One reason why this might be expected is if the dwarf galaxies, and
the clouds, were both representative of a single spectrum of systems
that inhabit the neighborhood of giant field galaxies.

It is reasonable to propose that if there is a continuity between the
nature of low, and high--redshift Lyman alpha absorption systems, then
the large population of HI clouds at high--redshift may likewise have
been loosely clustered about protogalaxies, and thus have been subject
to accretion to the more dominant gravitational potential in their
midst.  At modest redshifts ($z \sim 0.5$), Chen \etal (1998) has
found that clouds of Ly$\alpha$ equivalent width $ W \approx 0.3 {\rm
\AA}$ lie at a characteristic distance of $160 \hkpc$ ($h \equiv
H_{o}/100 \kmsMpc $), consistent with more local data (Lanzetta \etal
1995; Tripp \etal 1998).  It does not seem reasonable that these
clouds are rotationally supported, for the velocity dispersion of
galaxies within $\sim 5 \hMpc $ of the Local Group is on the order of
only $50 - 60 \kms$ (Giraud 1986; Schlegel \etal 1994), so that clouds
of extragalactic origin with tangential velocities of $\sim 200 \kms $
would seem to be highly unlikely. For clouds not rotationally
supported, however, the infall time is only $\sim 1/4$ that of the
look--back time to $z \sim 0.5$.  Therefore I conclude that these
clouds, and those at higher redshifts, must be being accreted to the
galaxies about which they cluster, and further, the distribution must
be being replenished to some significant degree since roughly the same
distributions maintain at $z
\approx 0.6$ as locally, as we have seen.

At redshifts in the range of 2 to 3, Fernandez-Soto \etal (1996) have
shown that there is strong clustering within $250 \kms$ among clouds
that have associated metal lines, while in the range, $2.7
\leq z \leq 3.7$, Chernormordik (1995) has found weak cloud clustering
within velocity scales of $\sim 150 \kms$ for clouds with columns
$\rm{N} \ge 10^{14} \cmtw$.  Using cross correlations of absorption
systems in double and group quasars at redshifts of $\sim 2$, Crotts
\& Fang (1989) found that, Lyman $\alpha$ systems with $W{o} \geq 0.4
{\rm \AA}$ cluster on scales as large as $0.7 \, (1.2) \hMpc$ for $\q0
= 0.5 \, (0.0)$, though there is an apparent significant variation
with environment.

These findings are broadly consistent with the physical association of
clouds with galaxies seen at low redshift, and we may therefore
postulate that high--redshift clouds may accrete to their more massive
neighbors on time--scales of $\sim 1^{+} \Gyr$, their dissipated
remains precipitating bursts of star formation.

\subsection{The Density of GCs and Clouds}

If Lyman $\alpha$ clouds are to cause the formation of GCs at
high--redshift, then the comoving density of Lyman alpha clouds at
high--redshift should be comparable to the density of GCs locally.
The density of GCs can be calculated from the luminosity function and
the luminosity--weighted specific globular cluster frequency, ${\rm
S_{N}}\equiv N_{t}/L_{15} $ (Harris \& van den Bergh 1981), where
$N_t$ is the number of GCs in the galaxy, and $L_{15}$ is the
luminosity in units of a galaxy with $M_V = -15$.  The number density
of GCs is given by the integral of the median specific GC frequency
times the luminosity--weighted Schechter (1976) function.  Given the
parameterization, $X=L/L^*$, we find the comoving density of GCs is,
\begin{equation}
n_{gc} = \langle S_{N} \rangle L^*_{15} \int X \Phi(X) dX \, \approx
\, \langle S_{N}\rangle L^*_{15} {\Phi}^*,
\end{equation}
where the result follows from setting $\alpha = -1$.  I make the
following substitutions: For a local sample of galaxies
(central--cluster ellipticals excluded), $ \langle S_{N}
\rangle \approx 4 $ (Harris 1991), $L_{15}^{*}
\approx 100$ if $M_{V}^{*} \approx -20.0$, and $\Phi^*
\simeq 0.014 \, \hthrMpc$ (Loveday \etal 1992).  Thus, 
\begin{equation}
n_{gc} \approx 5.5 \, \rm{h^3 \, Mpc^{-3}}.
\end{equation}

For clouds at high--redshift, the attributed comoving density of
clouds of radius $ r_{cl}$, and line density $dN/dz$ is, $ n_{cl} =
\frac{dN}{dz} \left(\frac{dz/dR}{\pi r_{cl}^2 (1 + z)^3} \right)$.  I
assume a low $\Omega$ FRW cosmology ($\Lambda = 0$), for which $dz/dR
\approx \frac{H_0}{c} (1 + z)^2$, where $H_0$ is the Hubble constant.
This yields,
\begin{equation}
n_{cl} \approx \frac{dN/dz}{\pi R_{o} r_{cl}^2 (1 + z)},
\end{equation}
where $R_o = c/H_o$.  A closed model doubles the expected density when $z=3$.

We must decide what fiducial value to substitute for $r_{cl}$.  A
numerical simulation of high--redshift self--shielded primordial
clouds, in which cold, isothermally distributed DM comprises $90 \%$
of cloud mass (Manning 1992), showed that the clouds responsible for
most absorption lines were large, just short of the size at which
thermal instability from self--shielding would occur in the central
regions.  These clouds were found to have H I columns of ${\rm Log} \,
N=14 \, (13)$ at projected radii of $ \sim 20 \, (45) \,
\hkpc $, and core radii on the order of $2 \kpc$, and core baryonic
mass of $\sim 1 - 4 \times 10^6 \msun$).  They have idealized thermally
broadened velocity dispersions in the range of $\sigma \approx 17-22
\kms$.  Meanwhile, Shull \etal (1996) have estimated the radii of local clouds
to be $\sim 100 \kpc $ at columns of $N \sim 10^{13} \cmtw $, and
Blitz \etal (1998) have found the upper limit cloud radius,
constrained by the tidal field of the Local Group, to be $\simeq 25$
kpc at columns of $N \simeq 10^{19} \cmtw$.  While these figures may
make the simulated cloud sizes at high--redshift seem low, the
50--fold drop in the far--UV metagalactic flux at the Lyman limit from
its value at $z=2.5$ (Haardt \& Madau 1996) will help to account for
this apparent disparity in absorption cross section.  In a study of
the lensed quasar $Q2345+007$A,B, Foltz \etal (1984) have placed a
lower limit on the characteristic $diameters$ of clouds at $\langle z
\rangle = 1.95$ of $5 - 25$ kpc (the uncertainty stems from the
unknown redshift of the lens).  These values simply reflect the range
of possible distances between beams.  However, a Monte Carlo analysis
of the correlated absorption systems in the spectra of Q$1343+2640$A,
B (Dinshaw \etal 1994), which has a similar ratio of ``hits'' to
``misses'' (h/m$\simeq 2$), found an inferred cloud diameter 2.5 times
that of the median beam separation.  If this correction is applied to
the Foltz \etal (1984) data, the cloud range in $radii$ becomes $7.5 -
62.5 \hkpc$.  However, recent work correlating absorption systems of
double and group quasars (e.g., Fang \etal 1996; Crotts \& Fang 1998)
find much larger values (typically few$\times$100s of kpc), but it is
not clear that these are large individual clouds rather than the
larger--scale clustering of much smaller individual clouds that we
have been discussing, which could be expected to give ``cloud
diameters'' of from $\sim 300 - 500 h^{-1}$ kpc, depending on the
equivalent widths probed and the method of analysis -- consistent with
their results.

For high redshift clouds ($z \sim 2-3$) I adopt a fiducial radius of
25 $\hkpc$ for individual H I clouds at columns of ${\rm Log} \, N
\simeq 14$.  A line density of $d{\rm N}/d {\rm z} \approx 100$ at a
redshift of $z\sim 3$ is assumed, based on data normalized to an
equivalent width limit of $W_o = 0.24 { \rm\AA }$ (Weymann \etal
1998), which, for $\rm T \approx 2 \times 10^4 $K, is quite close to a
column density, $N=10^{14} \cmtw $.  Substituting these values into
Eq. 3, we find,
\begin{equation}
n_{cl}(z=3) \approx 4.2 \,  \rm{h^3 \, Mpc^{-3}}.       
\end{equation}
For $\Omega = 1$, $n_{cl}=8.4 \, \rm{h^3 \, Mpc^{-3}}$.  Comparison of
Eqs. 2 and 4 shows $n_{gc} \approx n_{cl}$.  If there is a $spectrum$
of cloud sizes (I find the largest dominated the absorption
line--density), then since cloud number density, $n_{cl} \propto
r^{-2}$ (Eq. 3), the total number density of clouds may be
significantly larger, which may help the numbers to look better for
open cosmologies.  Also, if replenishment is occurring at
high--redshift by the condensation of clouds in voids and subsequent
movement toward neighboring galaxies, then the total number of
potential impactors could be substantially greater than estimated
above.

\section{The Physical Requirements for Cloud Survival}

For a modest sized protogalaxy, the infall velocities of an
intergalactic cloud can be expected to be in excess of 60 km/sec.  In
order to form a cluster, therefore, some small fraction of the
plunging cloud must survive the supersonic shock in a compact form.
Numerical simulations of homogeneous clouds subjected to interstellar
shocks have found that clouds are destroyed on time scales of a sound
crossing time (\eg, Klein, McKee \& Colella 1994; Murakami
\& Ikeuchi 1993).  However, numerical studies of the survival of
intergalactic, cold dark matter--held, ``mini--halo'' clouds
(Rees 1986) subjected to supersonic flows (Murakami \& Ikeuchi
1994) have shown that the cores of clouds confined by a dark halo may
survive extended periods of exposure to supersonic wind as long as the
central cloud density exceeds that of the ambient medium through which
it passes.  In the context of accretion to a galaxy, this will require
that the central density of the cloud should increase in time in order
that it may survive the increasing densities it will encounter as it
falls inward.  This in turn will require cooling.  For a stable
compression process, therefore, the cooling time scale, $\tau_c$, must
remain smaller than the time--scale for the increase in the ambient
protogalaxy gas density, $\tau_{d}$, at each stage of the cloud's
journey.  The timescale for change of density is expressible by the
equation,
\begin{equation}
\tau_d = \left(\frac{1}{\rho} \frac{\partial{\rho}}{\partial{t}} \right)^{-1}.
\end{equation}
I presume, for the sake of simplicity, that the gas and the DM are
distributed in an isothermal profile with density law,
$\rho={\cal{K}}/(4 \pi R^2)$, where $R$ is the galactocentric
distance, and ${\cal{K}}$ is a unit of mass per unit length.  ${\cal
K}$ will be referred to as the system mass--distribution constant.
For an $L^*$ galaxy, ${\cal{K}}={\cal{K}}^*\approx 1.14 \times 10^{7}
\Msun \pc^{-1}$.  I also presume that the distribution is truncated at
a distance $R_t$.  Then for $ R < R_t$, $M(R) = {\cal{K}} R$, and for
$R > R_t$, $M(R) = {\cal{K}} R_t$.  Manipulating Eq. 5, we find,
\begin{equation}
\tau_d = \frac{R}{2 v_i}, 
\end{equation}
where $v_i$ is the infall velocity.  With the assumed mass
distribution, we can calculate the potential:
\begin{equation} 
\Phi(R < R_t) = -G {\cal{K}} \left(ln \left(\frac{R_t}{R} \right) +
\left(1-\frac{R_t}{R_0} \right) \right), 
\end{equation}
where $R_0$ is the location, presumed in this case to be $\sim 2$ {\rm
Mpc}, of the turn--around radius, and at which the potential is set to
zero.  $R_t$ is assumed to be 500 kpc for an $L^*$--type object.
Conservation of energy requires that the infall velocity be given by
the equation, $v_{i}(R < R_t) \leq \sqrt{-2 \Phi(R < R_t)}$
the inequality stemming from dissipative effects.  Substituting this
into Eq. 6, and using the value, $\cal{K}^*$ (above), we find that the
timescale for change of density is,
\begin{equation}
\tau_d \gtrsim 1.38 \times 10^8 \left( \frac{R_{kpc}}{160} \right) \, \yr,
\end{equation}
where $R_{kpc}$ is the galactocentric radius in kiloparsecs.  At a
distance of $15$ kpc $\tau_d \approx 1.3 \times 10^7$ yr.

Dissipation is expected to begin in earnest at an estimated
galactocentric distance of $\sim 160 \hkpc $, for that is the
projected distance at which lines of sight encounter EWs of $\sim 0.3
\, \AA$, or columns of $\sim 10^{14} \cmtw$ (Chen \etal 1998).  It
will be important that the rate of cooling keep pace with the
time--scale for compaction.  I assume that dissipation balances the
rate of change the cloud potential energy so that the velocity remains
at a constant $50 \kms$. Thus we have,
\begin{equation}
v \frac{G M}{R^2} \int_{V} {\rho}_{cl} d V = n^2 k T \beta_B \int_{V} d V,
\end{equation}
where we will insert values for an $L^*$ galaxy, and cloud baryonic
density, $ {\rho}_{cl} = {n} \mu m_H $.  The case B recombination
cooling coefficient may be expressed as, $\beta_B = 9.17 \times
10^{-14} (T/20000)^{0.5} \, \rm{cm^3 \, s^{-1}} $.  From this relation
and Eq. 9, we derive,
\begin{equation}
T = \frac{1.127 \times 10^4}{n^{2/3}} \left(\frac {15}{R_{kpc}}
\right) \rm{K}.
\end{equation}
The cooling time--scale of the cloud can be defined,
\begin{equation}
\tau_{c} = \frac{\frac{3}{2} n k T}{\Lambda}.
\end{equation}
With $\Lambda=n^2 k T \beta_B$, we find
\begin{equation}
\tau_c \approx \left(\frac{2.313 \times 10^{15}}{n T^{1/2}} \right) \rm{s}.
\end{equation}
Combining Eqs. 10 and 12  we find,
\begin{equation}
\tau_c = \frac{6.9 \times 10^5}{n^{2/3}} \left( \frac{R_{kpc}}{15}
 \right)^{\frac{1}{2}} \yr.
\end{equation}
The pressures behind the head of the bow shock will result in baryon
densities $\rho_{cl} \approx 4 {\cal{M}}^2 \rho_{gal}(R)
(T_{gal}/T_{cl})$ during the compressonal stage, where ${\cal{M}}$ is
the Mach number, expected to range up to $\approx 10$.  Assuming an
$L^*$ galaxy, at $R \approx 15 \kpc$, and assuming $90 \%$ DM, then
the galactic baryon number density is, $n_{gal} \gtrsim 10^{-2}$.
Thus, we may expect $n \gtrsim 1$, and should eventually reach
densities of order $10^4 \, \cmthr$.  This will assure that $\tau_{c}$
is much less than $\tau_d$.

For galactocentric radii of interest, the cooling time scale within
the core of the cloud appears to be comfortably less than $\tau_d$.
During the final stages of accretion, another
term must be added to Eq. 9 to represent the deceleration of the
cloud.  This will compensate for the decline of the first term,
leading to only modest changes in the derived cooling time--scale.  It
must be emphasized that there \emph{should} be a comfortable gap
between $\tau_d$ and $\tau_c$, for we don't expect real galaxies to
have entirely smooth density profiles.

\section{The Mechanism for Cloud Compaction}

In order to satisfy the requirements that these events produce
globular clusters, the cloud must be compressed to densities
comparable to, or in excess of that found in globular cluster cores.
Powered by the kinetic energy of the cloud, the combination of
pressurization and radiative cooling is capable of compacting the
cloud.  In this section I discuss three important facets of the GC
formation scenario.

\subsection{Dynamical Features of the Interaction}

Let us now suppose that a centrally condensed cloud (core $\sim 3-5
\kpc$ diameter, is drawn into the potential well of a gas--rich 
protogalaxy.  When the cloud begins to encounter the dissipated
gaseous halo at velocities of $\sim$Mach 10, a bow shock is
established, forming a shocked shell about the cloud.  In the frame of
the shock, galactic gas is quickly decelerated by the shock within
this shell, imparting a momentum to the infalling cloud.  Only gas of
density comparable to that of the wind is stripped--- a result of the
small mean free path in relation to cloud size --- while the dense
core is found to survive in excess of $3 \gyr$ (Murakami \& Ikeuchi
1994).  Accordingly, we should expect only the core to survive intact,
carrying $\sim \rm{few} \times 10^6 \msun$ of baryons, as noted in
\S2.1,  in the range of $10^{-2} - 5 \times 10^{-4}$ of the total cloud
baryonic mass.  The shocked and stripped remainder, much of which
would be caught up in eddies, may participate in star formation in the
manner suggested by Binney (1976), contributing to the halo population
at relatively low efficiencies.  The remnants of this would settle
toward the disk.

While the DM component is thought to play an important role in
containing the cloud in the IGM (Blitz \etal 1998; Shull \etal 1996),
once the ram pressure, integrated over the surface of the cloud,
exceeds the maximum force between the baryonic cloud and the DM cloud,
the DM halo would be lost.  This is expected to occur in the early
stages of intense dissipation, at distances less than $\sim 160
\hkpc$ (see \S3).  When the deceleration on the cloud begins to 
dwarf its self--gravity, the cloud will shift and compress to
approximate an exponential atmosphere with a scale--height determined
by the deceleration rate, which in turn will heat the cloud.

Within this environment, two forces act upon the cloud which tend to
induce vorticity, a shift of the higher density gas within the cloud
toward the shock front, displacing lower density gas toward the side,
and the shear force due to the stripping flow inside the bow shock.
While by itself, molecular viscosity is not sufficient to appreciably
effect the vorticity, a turbulent viscosity will be produced.  The
shock will stand off from the cloud due to the high pressures
immediately behind the shock, but the dense, cool cloud immediately
``downstream'' of this region will be subject to Rayleigh--Taylor
instabilities since cloud deceleration will result in denser material
overlying less dense material.  As these perturbations grow they will
become subject to the shear flow closer to the shock, transforming
them into Kelvin-Helmholtz (K--H) instabilities.  These instabilities
would continue to grow along the ``fetch'' of the interface of
galactic and cloud gas, causing a turbulent viscosity and an increased
transferral of angular momentum to the cloud.  The combined effect of
these forces would be to induce vorticity into a torroidal volume
whose axis is aligned with the vector of the cloud's motion.  During
the stage of intense dissipation, it is better to compare it to an
inverted convection cell than to a vortex ring, for unlike the vortex
ring, the cloud is contained from without by the shock and by the
deceleration, rather than from within by a vacuum.

During the final stage of intense dissipation, the increased
deceleration means that gas which has been stripped need not be
irrevocably lost to the cloud.  Two factors are responsible: 1)
stripped material will be pushed toward the central axis of motion at
$\sim$sonic speeds because of the near vacuum behind the cloud, 2) the
deceleration of the cloud will result in a relative (negative)
acceleration between the stripped material and the shock front.  The
re--accretion of this gas to the cloud would reinforce the pattern of
motion established in the early adjustment of the cloud.  It is an
important feature of this scenario that the stripped, and eventually
re-incorporated material will contain metals present in the
protogalaxy, which have been mixed with cloud gas by K--H
instabilities.  The well--established observational constraint that
the metallicity in almost all GCs is uniform to within 0.05 dex
(Suntzeff 1993), requires that turbulent mixing should occur well
before star formation begins (Murray and Lin 1993).  In this scenario,
the enhancement of the metallicity, and the mixing, occur as a result
of the natural dissipative process of accretion outlined above,
comfortably preceding fragmentation (see \S4.4).

\subsection{The Stability of the Plunging Cloud}

The modeling of intergalactic clouds held by DM (see \S2.1) indicates
they may be centrally condensed.  Furthermore, centrally condensed
clouds have been found to survive shocks (Murakami \& Ikeuchi 1994).
Therefore, if this model of \Lya clouds is correct, they should be
expected to survive the gentle onset of ram pressure.  The vorticity
induced by the shear flow inside the shock does not appear to be a
destabilizing influence,  because the cloud is contained by the shock
shell and by deceleration, as described above.  However, when the
cloud decelerates to subsonic velocities, it will evolve toward
an ordinary, albiet large, vortex ring.  Here, the orderliness of the
motion will be critical in determining the length of time the cell
should survive.  Studies of vortex rings at relatively high Reynolds
numbers show that sinusoidal bending modes may grow, and eventually
destroy the ring (Widnall \& Sullivan 1974).  However, these modes,
whose number are few when vorticity is widely distributed, would be
damped by the shock which wraps around the cell during the period of
intense dissipation, but may grow after it is no longer supersonic.
At the very minimum, the cell will survive for an eddy turn--over
time, given that the ordered motion will produce a vacuum in a ring
which will cause the self--induced motion characteristic of vortex
rings.  For a cell of dimension $10$ pc, the eddy turn--over time is,
\begin{equation}
\tau_e \approx 7.6 \times 10^5 \left(\frac{c_s}{v} \right) \, \yr,
\end{equation}
where $v$ is the peak tangential velocity of the vortex cell.  The
self-inducing motion of the vortex ring is likely to cause it to last
many times greater than this.  However, accurate estimates of this
must await a careful numerical simulation, as energy and time--scale
arguments are held hostage to uncertainties in the levels of turbulent
viscosity during the stage of intense dissipation.

\subsection{Energetic Requirements for Cloud Compression}

If radiative cooling can keep pace with heating, we may assume
isothermal pressurization.  If the cloud core has a baryonic mass of
$m_{cl}$, an initial baryonic density of $\rho_{i}$, and a final
density of $\rho_{f}$, the required work for compression is,
\begin{equation}
W = m_{cl} c_s^{2} \, ln\biggl(\frac{\rho_{f}}{\rho_{i}}\biggr).
\end{equation}
Values for $W$ are insensitive to variations by factors of a few in
the initial or final densities.  Equating this work to the kinetic
energy of the cloud core we find that the required infall velocity
$v_{i}$, is given by,
\begin{equation}
v_i = c_s \sqrt{2 \, ln(\rho_{f}/\rho_{i})},
\end{equation}
where $c_s$ is the sound speed of the cloud core.  For a final density
equal to that of the median Galactic globular cluster core densitiy,
$\sim 10^{3} \Msun {\rm pc}^{-3}$ (Djorgovski 1993), a factor of
compression in the range of $\sim10^{6-8}$ is expected, and yields a
required infall velocity, ${v_{i}/c_s} \approx 5.0 - 6.0 $.  For a
temperature of $2.0 \times 10^4$ K, the minimum required infall
velocity is $\sim 60 -75 $ km/s, a velocity attainable even in dwarf
systems (see \eg, Wyse \& Silk 1985).

\subsection{The Transformation of Cloud to Cluster}

The final stage of the transformation is reached when the velocity
approaches subsonic levels.  During the supersonic period of
dissipation we expect that the cloud size will be decreasing -- not so
much due to stripping, but because of its compaction.  When the ram
pressure, $\rho v^2$, begins to decline, the cell will gradually
transform from one contained by the shock shell to one contained by
its own vorticity.  This will be occassioned by some expansion, and as
it expands, the cloud deceleration will increase, leading to a
relatively rapid change of state.  Though the density of the
surrounding medium should by this time be rather large, it is likely
--- indeed it is imperative --- that the density in the central axis
of the vortex ring should be orders of magnitude greater.  The
disappearance of the shock will allow the rapid cooling of cloud gas
at the head of the shock, where it is densest.  Yet the pressure on
this gas is sustained by pressures from all directions: the gas near
the axis is subjected to an axially symmetric centrifugal force from
the vortex ring; toward the host galaxy there remains the ram pressure
of the cloud at near sonic velocities; and finally, pressure is
exerted by material which has been stripped, but is re--joining its
cloud, now falling with increased velocity due to the increased
deceleration.

To assess the plausibility and efficiency of star formation at this
point we need to know the collapse time--scale for gas at the
expected densities, the transit--time scale for gas in the axis of the
vortex, and some notion of the duration of the vortex cell, which is
responsible for maintaining the radial pressure along the axis of the
cell.  The time--scale for cloud collapse is,
\begin{equation}
\tau_{coll} \approx1/\sqrt{G \rho} \sim 4.7 \times 10^5 \yr,
\end{equation} 
for the central density cited in \S4.3.  The time--scale for transit
down an axis of $10$ pc is $\sim \tau_e$, given by Eq. 14, where, by
assumption, $v < c_s$.  As the cell is slowed, this velocity will
fall, in time providing a sure opportunity for gravitational collapse.
Note that the cooling time--scale (Eq. 13) for baryon densities of
order the central GC densities ($n_{cl} \approx 4 \times 10^4 \cmthr
$) is well within these time--scales.  The cell itself should be
stable for well over an eddy turn--over time--scale (Eq. 14), in
keeping with the large durations of more modest--sized vortex rings
which may travel distances that are many times their diameters
(Widnall \& Sullivan 1973) before disruptionl.  A more severe threat
to the survival of the vortex cell is condensation of gas and star
formation, but by then, we don't care.

Star formation rates, and efficiencies are now of concern.  In a study
of 97 normal, and star--forming galaxies, Kennicutt (1977) found that
the disk--averaged star formation rates were found to fit to a Schmidt
(1959) law with index $N = 1.4 \pm 0.15$. It was found that,
\begin{equation}
\Sigma_{SFR}=(2.5 \pm 0.7) \times 10^{-4} \left(\frac{\Sigma_{gas}}{1
\Msun \rm{pc}^{-2}} \right)^{1.4 \pm 0.15} \, \Msun {\yr}^{-1} {\kpc}^{-2}.
\end{equation}
If we imagine gas with a density equal to peak central GC densities
(\S4.3), distributed over a 10 pc cube, then the indicative star
formation density is $\sim 0.25 \Msun {\yr}^{-1} {\pc}^{-2}$, and
would yield $1.25 \times 10^7 \, \Msun$ within a plausible time-scale
of $5 \times 10^5 \, \yr$, a mass larger than the $10^6 \, \Msun$
contained in the cube.  Star formation efficiency would vary with
index $N-1$, or, $\Sigma_{gas}^{0.4}$.  At projected final cloud
densities, this implies star formation efficiencies $40$ times greater
than that of a normal disk with a gas surface density of $\sim 1 \,
\Msun \rm{pc}^{-2}$. This would rival that of central starbursts (Kennicutt
1977).

\section{Implications for Field Galaxy Formation}

A model in which galaxies are subjected to periodic accretion of low
metallicity clouds is consistent with the ``chaotic'' scenario of
galaxy formation of Searle \& Zinn (1978), and Searle (1980), now
often referred to as hierarchical structure formation.  The GC
formation scenario requires a picture in which galaxy growth is a
protracted process.  That the accretors are probably DM--held, and
generally relaxed, seems required by the specific scenario developed
here.  Shull \etal (1996) estimate that these clouds have a total mass
(baryons plus DM) of $M_{tot}=10^{9.8} \Msun {\rm R}_{100}{\rm
T}_{4.3}$, where $R_{100}$ is the radius in units of 100 kpc at which
the column density of neutral hydrogen is $10^{13}
\cmtw$ , and $T$ is in units of $10^{4.3} {\rm K}$.  This value is
consistent with a mass distribution constant, ${\cal K} \approx10^{5}
\Msun {\pc}^{-1}$, with a circular velocity $v_c \sim 20 \kms$, as
observed by Blitz \etal (1998).  Somewhat smaller clouds may also be
realistic.

The accretion of such a cloud would produce great turbulence.  Might
there be low--redshift counterparts of proto--late--type galaxies?
High resolution VLA observations of 5 actively star forming blue
compact dwarfs (BCDs) reveal kinematically distinct clumps of H I, and
turbulent outer H I envelopes (on scales of a few$\times$100 pc).
This is broadly suggestive of the accretion of H I clouds.  That the
median gas depletion time scale is on the order of 1 $\Gyr$ also
argues for extragalactic replenishment.  Thus these young galaxies may
resemble our Galaxy in its earliest formative stages.  It is easily
seen that the accretion of one or two hundred clouds of $\sim 10^{9.8}
\msun$ can account for the total mass of an $L^*$ galaxy within
$R_{gc} \approx 100 \kpc$.

While the formation of GCs is somewhat incidental to the
process of galaxy formation as outlined above, yet they remain
potentially sensitive probes of the state of the Galaxy at its early
formative stages.  The GC formation scenario suggests that among the
requirements for cluster formation are two constraints on the
(proto)galaxy -- that it must have a large gas column density, and a
mass distribution constant large enough to support required circular
velocities of $\gtrsim 40
\kms$, a value somewhat below that characteristic of dwarf spiral
galaxies.  The former follows by the observation that ram pressure
must slow the cloud to subsonic velocities: $\int \rho_{gal} v^2 S(t)
d \, t \approx - \int m_{cl}(t) \frac{d \Phi(t)}{d \, R} d \, t $,
where $\rho_{gal}$ is the galaxy baryon density, $S(t)$ is the cloud
surface area, and stripping will reduce the cloud baryonic mass,
$m_{cl}$, in time.  Alternatively,
\begin{displaymath}
\int_{R_o}^{R}  \rho_{gal}(R) v(R) S(R) d\, R \approx - \int_{R_o}^R \frac{G
M(R) m_{cl}(R)}{R^2} d \, R.
\end{displaymath}
A detailed numerical simulation will be
required to properly evaluate these integrals, but clearly, if the
column of galactic gas, $\int \rho(R) \pi r_{cl}^2(R) dR $ is too
small, then the equation cannot be met.  The mass--distribution
requirement, together with the equation, $V_{rot}=\sqrt{G {\cal{K}}}$,
is essentially that ${\cal{K}} \gtrsim 1.2 \times 10^6
\Msun \rm{pc^{-1}}$.  This threshold for
GC formation results from the observation that an infall velocity of
$60 \, \kms$, which is at the lower end of the range satisfying the
energetic requirements for cloud compression, is $\sim \sqrt{2}$ times
the circular velocity at that distance.  Once the galaxy mass is great
enough, GCs would be formed.  However, in response to an increased
specific star formation rate accompanying the formation of the bulge
and the disk, the gas column should decrease, signaling an end of the
halo and GC formation epoch.  During the stage of this retreat, we
would expect that, if it relaxed into a thick, slowly rotating disk,
then GCs would have a lower probability of being formed if the cloud
were plunging normal to the disk, due to the lower gas columns.  Thus,
the metal--rich ``disk'' globulars, with their flattened distribution
(Armandroff 1993), and the thick disk itself, may be evidence favoring
this interpretation.

\section{Implications for Cosmology}

While the redshifts of formation implied by this scenario may strike
the reader as quite low, recent findings, largely based on
\emph{Hipparcos} data, the inclusion of helium diffusion into the
cores of stars, and an improved equation of state, yield new distance
determinations based on the main sequence turnoff, which in turn imply
that the oldest globular clusters may be much closer to $12 \gyr$ old.
This is a $\sim 20 \%$ reduction from pre-{\emph{Hipparcos}} values
(for a recent review, see Chaboyer 1998).  If we allow that these
oldest GCs were formed at $z=3$, then given this GC age, we may derive
$H_o$ as a function of cosmology.  Accordingly, we find that if
$\Omega=1$, we would require $H_{o}=47.7 \, \kmsMpc $.  If $\Omega=0$,
$H_{o} = 61.3
\, \kmsMpc $.  For a flat $\Lambda$ model with $\lambda=0.7$, $H_{o} 
= 67.8 \, \kmsMpc$.  If the formation epoch instead were at $z \approx
4$, then the attributed Hubble constant is increased by less than $6.5
\%$ for all three models.

It has been reported that GCs may have a significant range of ages.
Recent estimates are in the range of $\sim 5 \gyr$ (Fusi-Pecci
\etal 1995; Chaboyer, Demarque \& Sarajedini 1996).  Using 
standard astrophysical formulae, I find that the range of time between
the plausible redshift range $z = 1.5 - 3$, the expected range of GC
formation redshifts, is, 1.79, 2.44 and 2.20 $\gyr$ respectively for the
$\Omega=1$, 0, and $\Lambda=0.7$ cosmologies, where for
self--consistency I have used the specific value of $H_o$ derived for
each case.  These ranges are smaller than the observed value
cited above, but not so much as to cause alarm at this early stage.

\section{Conclusion}

High resolution HST spectra at low redshift have disclosed the
existence of a surprisingly large population of Lyman $\alpha$ clouds,
most of which are thought to be clustered within 2 $\hsfMpc$ of bright
field galaxies.  The juxtaposition of evidence for cloud clustering,
and for major accretion events of low metallicity gas onto large field
galaxies, suggests a \emph{causal} relationship.  By following the
commonalities of cloud clustering among the low--redshift population
out to high--redshift, it becomes clear that clouds at high--redshift
certainly had the opportunity to accrete to the protogalaxies about
which they are thought to cluster.  The dramatic disappearance of
Lyman $\alpha$ clouds at high--redshifts provides the link between the
clustering of clouds and the heightened star formation rates in field
galaxies seen at $z \sim 1-3$ (Madau \etal 1998).  It is suggested
that the central regions of many of these clouds might plausibly have
been transformed into GCs, while the less--strongly held gas may have
contributed to the formation of the stellar halo.  It has been shown
that the energetics of cloud compression are favorable, as are the
numerical coincidences of the comoving densities of GCs, and clouds at
high--redshift.  The juxtaposition of the projected redshifts of
formation with new globular cluster ages implies values of $H_o$ that
are reasonable.  However, the predicted cosmology-specific age range
of GCs appears to be lower than recent work would imply.  This
interesting fact promises to be a goad to stimulate future work.

\bigskip

I would like to thank the anonymous referee for many comments and
suggestions that have improved this paper.  In addition, I thank Hyron
Spinrad, Daniel Stern, Christopher F. McKee, Robert Fisher, and Dean
McLaughlin for helpful suggestions during the planning and production
of this paper.

\end{document}